\documentclass[prb,aps,twocolumn,showpacs,superscriptaddress]{revtex4-1}
\usepackage{graphicx}
\usepackage{bm}
\usepackage{bm,color}

\newcommand{\be}{\begin{eqnarray}}
\newcommand{\ee}{\end{eqnarray}}

\begin{document}

%\preprint{APS/123-QED}

\title{
Phase diagrams of Bose-Hubbard model and antiferromagnetic spin-1/2 models \\
on a honeycomb lattice}
% Force line breaks with \\

\author{Takashi Nakafuji and Ikuo Ichinose}
% \altaffiliation[Also at ]{}%Lines break automatically or can be forced with \\
\affiliation{
Department of Applied Physics, Nagoya Institute of Technology, 
Nagoya, 466-8555, Japan}

\date{\today}% It is always \today, today,

\begin{abstract}
Motivated by the recent experimental realization of the Haldane model by ultracold 
fermions in an optical lattice, we investigate phase diagrams of the 
hard-core Bose-Hubbard model on a honeycomb lattice. 
This model is closely related with a spin-1/2 antiferromagnetic (AF) quantum spin
model.
Nearest-neighbor (NN) hopping amplitude is positive and it prefers an AF
configurations of phases of Bose-Einstein condensates.
On the other hand, an amplitude of the next-NN hopping depends on an angle variable
as in the Haldane model.
Phase diagrams are obtained by means of an extended path-integral Monte-Carlo
simulations.
Besides the AF state, a 120$^o$-order state, there appear other phases including
a Bose metal in which no long-range orders exist.
\end{abstract}

\maketitle
%%%%%%%%%%%%%%%%%%%%%%%%%%%%%%%%%%%%%%%%%%%%%%%%%%%%%%%%%%%%%%%%%
\section{Introduction}\label{Intro}

In the recent years, systems of ultracold atomic gases in optical lattices have
attracted much attention.
Since the systems have the high controllability and versatility,
they provide us with a quantum-simulation platform for studying 
strongly-correlated systems in condensed matter physics, lattice quantum
field theories, etc \cite{coldatoms}.
It is established nowadays that low-energy properties of
interacting Bose gases in optical lattices are described by the Bose-Hubbard model \cite{BHM}
and its extension.
The idea of quantum simulation by using ultracold atomic systems is also applied to 
theoretical models that have been regarded as only academic ones.
Theoretical predictions for such models are expected to be observed 
by experiments on ultracold atoms.
Recently, generation of artificial gauge fields in ultracold atomic systems
in optical lattices was succeeded in the experiments by rotating/shaking
optical lattices or by using laser-assisted tunneling \cite{mg_optical,Honeycomb_ex}.
These techniques enable an experimental realization of the Haldane model 
with ultracold fermions.
This model was introduced as a fermionic tight-binding model on a honeycomb
lattice that breaks time-reversal symmetry without a net magnetic flux, 
and it exhibits interesting topological properties as a result of 
the next-nearest-neighbor (NNN) complex hopping term \cite{Haldane}. 

In the present study, we consider a bosonic analog of the Haldane model, 
which is called Bose-Haldane-Hubbard model (BHHM).
A recent study reported its ground-state properties and low-energy excitations 
at unit filling \cite{BHM1}.
In particular, we are interested in the hard-core boson limit where the
on-site repulsive interaction $U$ is very large ($U\to \infty$).
We call this model hard-core boson Haldane-Hubbard model (hard-core BHHM).
In the previous works \cite{Kuno,Nakafuji}, we studied the dipolar hard-core BHHM 
by means of the extended Monte-Carlo (MC) simulation (see later section) and 
showed that the model has very rich phase diagrams.
In this paper, we study the case in which the nearest-neighbor (NN) 
hopping amplitude is positive and the NNN hopping is complex depending on 
an angle $\phi$.
Generally, hard-core boson models can be mapped onto spin-1/2 models.
The hard-core
BHHM is closely related to the quantum spin-1/2 models on the honeycomb lattice.
Positive hopping amplitudes in the boson model correspond to 
antiferromagnetic (AF) exchange couplings in the spin model.

Hard-core BHHM on a small lattice was studied by the exact diagonalization
methods \cite{varney}.
The result suggested that this model had a quantum-liquid state, named  
Bose metal (BM).
The BM corresponds to a gapless spin-liquid state.
However, the existence and nature 
of this state is not established yet. 
Therefore, it is interesting and also important to study the existence and 
nature of the BM in larger systems.

In this paper, we shall study the hard-core BHHM on the honeycomb lattice
by means of the extended path-integral MC simulations.
As stated previously, we take the NN hopping amplitude positive
in the present study.
In order to study the maximum frustrated case, we also introduce a tunable
phase in the NNN hopping such as $J_2e^{i\phi}$ with $J_2>0$.
We shall clarify the phase diagrams in the two-dimensional $(\phi-J_2)$ plane,
etc.
Obtained phase diagrams of the BHHM shed light on the properties of 
the spin-1/2 frustrated AF-$XY$ and AF-$XXZ$ models on the honeycomb lattice.
As we mostly consider the case of the half filling, various superfluid (SF) states
appear in the phase diagram.
Correlation of the phase degrees of freedom of the Bose-Einstein condensate (BEC)
has a specific pattern in each SF.
Therefore, the terminology of spin is useful to distinguish SFs, and we shall often
use it.

The present paper is organized as follows.
In Sec.~II, we introduce the hard-core BHHM and the path-integral techniques
using the slave-particle representation.
An effective model is derived by integrating fluctuations in local density
as in the previous works \cite{Kuno,Nakafuji}.
The derived model has a positive-definite action and the MC simulation
is applicable for it.
In Sec.~III, we explain the extended MC simulation by introducing a lattice
in the imaginary-time direction.
Section IV exhibits the results of the numerical study.
We first study the low-temperature ($T$) phase diagram of the BHHM with the vanishing
NN repulsion and $\phi=0$.
We show that for a small system, the obtained phase digram is in good agreement
with that obtained by the exact diagonalization for the same system 
size \cite{varney}.
However, we found that the phase diagrams have rather strong system-size
dependence.
Besides the expected spin-ordered states, there exists a state that seems to have
no long-range orders (LROs), and we call that state BM as in the previous work.
We also study the finite-$T$ phase transition.
The result indicates that the BM has no LROs.
In Sec.~V, we study the phase diagrams of the system with various $\phi$.
Introduction of a finite $\phi$ diminishes the frustration and stable phases
form.
Among them, a new phase that we call ${\bf k}=(-\pi/\sqrt{3}, 0)$ forms
between the AF and 120$^o$-order states.
Finally in Sec.~VI, we consider the effect of the NN repulsion, which corresponds
to the $z$-component AF coupling, $\sum S^z_iS^z_j$.
Charge density wave (CDW) forms as the NN repulsion is getting large.
Section VII is devoted for conclusion.

%%%%%%%%%%%%%%%%%%%%%%%%%%%%%%%%%%%%%%%%%%%%%%%%%%
\section{Model Hamiltonian and path-integral formulation}\label{Model}

The Hamiltonian of the hard-core BHHM on a honeycomb lattice is given as follows:
\begin{eqnarray}
&&H_{\rm BH}=H_0+H_{\rm NN}, \nonumber \\
&&
H_{0}=J_{1}\sum_{\langle i,j\rangle}(a^\dagger_{i}a_{j}+\mbox{H.c.})
+J_{2}\sum_{\langle\langle i,j\rangle\rangle}(e^{i\phi_{ij}}a^\dagger_{i}a_{j}+\mbox{H.c.}),
\nonumber  \\
&&H_{\rm NN}=V\sum_{\langle i, j\rangle}n_in_j,
\label{H0}
\end{eqnarray}
where $a^\dagger_{i} \ (a_{i})$ is the hard-core boson creation (annihilation) operator 
at site $i$ and $n_{i}=a^\dagger_{i}a_{i}$ is the corresponding number operator.
In this Hamiltonian, $J_1$ and $J_2$ are the NN and NNN hopping amplitudes, 
respectively, and $V(>0)$ is the parameter of the NN repulsion, which may be
produced by the dipole-dipole interaction \cite{DDI}.
The NNN hopping term is complex $J_2e^{i\phi_{ij}}$ with the phase $\phi_{ij}=\pm\phi$
(th sign $\pm$ will be specified in the later discussion), 
which is experimentally feasible by the time-periodic driving of the honeycomb 
optical lattice.

Reflecting the hard-core nature, the physical Hilbert space consists
of states in which the particle number at each site is less than unity. 
In order to incorporate the local constraint faithfully, we employ the following 
slave-particle representation,
\begin{eqnarray}
a_{i}=h^\dagger_{i}b_{i},
\label{slave}
\end{eqnarray}
with the constraint,
\begin{eqnarray}
(b^\dagger_{i}b_{i}+h^\dagger_{i}h_{i}-1)|\mbox{Phys}\rangle=0,
\label{constraint}
\end{eqnarray}
where $b^\dagger_{i}$ ($b_{i}$) and $h^\dagger_{i}$ ($h_{i}$) are the boson and hole operator at site $i$, respectively.
$|\mbox{Phys}\rangle$ denotes the physical subspace of the slave particles 
corresponding to the hard-core boson.
From Eqs.(\ref{slave}) and (\ref{constraint}), it is not difficult to show
that the operators $a_i$ and $a^\dagger_i$ on the same site satisfy the
anti-commutation relation such as 
$\{a_i, a^\dagger_i\}=1$ and $\{a_i, a_i\}=\{a^\dagger_i, a^\dagger_i\}=0$,
whereas the usual bosonic commutation relations such as 
$[a_i,a^\dagger_j]=0$, etc., for $i\neq j$.
For example in the slave-particle representation,
\begin{eqnarray}
(a_ia^\dagger_i+a^\dagger_ia_i)b^\dagger_i|0\rangle
&=&(h^\dagger_ib_ib^\dagger_ih_i+b^\dagger_ih_ih^\dagger_ib_i)b^\dagger_i|0\rangle
\nonumber \\
&=&b^\dagger_i|0\rangle,
\nonumber
\end{eqnarray}
where we have used the ordinary bosonic commutation relations of the
slave particles $b_i$ and $h_i$ and the constraint Eq.(\ref{constraint}).
The standard path integral for the Bose particles 
{\em with the faithful local constraint} guarantees the above 
hard-core commutation relations.

In most of the later discussions, we consider the half-filling case,
which corresponds to the case $\langle \sum_i S^z _i\rangle=0$ in the spin 
system.

Note that the system described by the Hamiltonian [Eq.(\ref{H0})] is closely
related to a $s=1/2$ AF spin model on the honeycomb lattice by
the correspondence such as
$a^\dagger_{i} \to S^{+}_{i}$, $a_{i} \to S^{-}_{i}$, $(n_i-{1 \over 2}) \to S^z_i$,
$H_{\rm BH}\to H^{\rm S}$,
\begin{eqnarray}
H^{\rm S}&=&H^{\rm S}_0+H^{\rm S}_{\rm NN}+H^{\rm Z}, \nonumber  \\
H^{\rm S}_{0}&=&J_{1}\sum_{\langle i,j\rangle}(S^{+}_{i}S^{-}_{j}+\mbox{H.c.})
\nonumber \\
&&+J_{2}\sum_{\langle\langle i,j\rangle\rangle}(e^{i\phi_{ij}}S^{+}_{i}S^{-}_{j}+\mbox{H.c.}),
\label{spin-1/2XY} \\
H^{\rm S}_{\rm NN}&=&V\sum_{\langle i, j \rangle}S^z_iS^z_j, \;\;\;
H^{\rm Z}={3\over 2}V\sum_i S^z_i, \label{spin-z} 
\end{eqnarray}
where $S^{\pm}_{i}$ are the operators that flip a spin at site $i$.
$J_1$ and $J_2$ correspond to the NN and NNN spin exchange couplings.
For $J_1, \ J_{2} > 0$, the NNN coupling generates the frustration whose strength
is controlled by the parameter $\phi$ for $0\leq\phi<\pi$.
The chemical potential is introduced in the boson system in order to cancel 
the Zeeman term $H^{\rm Z}$ although we do not show it explicitly.

In our previous numerical studies \cite{EMC,Kuno,Nakafuji}, 
the results of the MC simulations
show that density fluctuation at each lattice site is not large even in
the spatially inhomogeneous states like a density-wave state.
From this observation, we expect that the following term appears,
\begin{eqnarray}
H_{1} &=& V_{0}\sum_i\Big(
(b^\dagger_{i}b_{i}-\rho_{bi})^2+(h^\dagger_{i}h_{i}-\rho_{hi})^2\Big),
\label{H1}
\end{eqnarray}
where $\rho_{bi}$ ($\rho_{hi}$) is the parameter that controls the mean
density of boson (hole) at site $i$, and $V_{0} (>0)$ controls its fluctuation from 
the mean value $\rho_{bi}$ ($\rho_{hi}$).
We impose the local constraint such as $\rho_{bi}+\rho_{hi}=1$ in the MC
simulation.
It is expected that the hopping terms in $H_{\rm BH}$ (i.e., $H_0$) enhance
homogeneous configurations, and induces terms such as $H_1$.
In Ref.~\cite{EMC}, we discussed how $H_1$ [Eq.~(\ref{H1})] appears from the 
hopping terms in the Hamiltonian and 
the rough estimation of the parameter $V_0$ gives 
$V_0 \sim J_1/\rho_{bi}$.
However, the precise value of $V_0$ depends on the dynamics of the phase degrees 
of freedom of the slave particles \cite{EMC}.
Then in the present work, we put typical values for $V_0$ and verify the
stability of the numerical results \cite{slaveMC}. 
We explicitly add this term to the Hamiltonian and consider the system 
$H_{\cal T}=H_{\rm BH}+H_{1}$.
The existence of $H_{1}$ is very useful for study of the quantum many-particle systems by the path-integral MC simulation.

The model $H_{\cal T}$ is studied by the path-integral methods \cite{EMC}.
To this end, we introduce the coherent states for the slave particles as follows,
\begin{eqnarray}
&& b_i|\varphi_{bi}\rangle=\varphi_{bi}|\varphi_{bi}\rangle=\sqrt{\rho_{bi}+\delta\rho_{bi}} 
\ e^{i\theta_{bi}}|\varphi_{bi}\rangle, \nonumber \\
&& h_i|\varphi_{hi}\rangle=\varphi_{hi}|\varphi_{hi}\rangle=\sqrt{\rho_{hi}+\delta\rho_{hi}} 
\ e^{i\theta_{hi}}|\varphi_{hi}\rangle,
\label{coherent}
\end{eqnarray}
where $\delta\rho_{bi}$ ($\delta\rho_{hi}$) is the quantum fluctuation of
the density around the mean value $\rho_{bi}$ ($\rho_{hi}$) at site $i$
and $\theta_{bi}$ ($\theta_{hi}$) is the phase degrees of freedom.
In the path-integral representation of the partition function $Z$, the action contains
the imaginary terms like $\int d\tau \bar{\varphi}_{i}(\tau)\partial_\tau \varphi_{i}(\tau)$,
where $\bar{\varphi}_{i}$ stands for the coherent field of 
$b^\dagger_{i}$ ($h^\dagger_{i}$) and 
$\tau$ is the imaginary time, i.e.,
\begin{eqnarray}
Z&=& \int [D\varphi_{b} D\varphi_{h}]
\exp\Big[-\int d\tau\Big(\bar{\varphi}_{bi}(\tau)\partial_\tau \varphi_{bi}(\tau) \nonumber \\
&&+\bar{\varphi}_{hi}(\tau)\partial_\tau \varphi_{hi}(\tau)+H_{\cal T}\Big)\Big],
\label{partition}
\end{eqnarray}
where $H_{\cal T}$ is expressed by the slave particles and the above path integral
is calculated under the constraint Eq.(\ref{constraint}).
For the existence $H_{1}$, we separate the path-integral variables $b_{i}$ and $h_{i}$
as Eq.(\ref{coherent}) and we integrate out the fluctuations $\delta\rho_{bi}$ 
and $\delta\rho_{hi}$.
However,
there exists the constraint such as $\delta\rho_{bi}+\delta\rho_{hi}=0$ on performing
the path-integral over $\delta\rho_{bi}$ and $\delta\rho_{hi}$.
This constraint can be readily incorporated by using a Lagrange multiplier $\lambda_i$,
\begin{eqnarray}
\prod_{\tau}\delta(\delta\rho_{bi}+\delta\rho_{hi})=\int d\lambda_i
e^{i\int d\tau (\delta\rho_{bi}+\delta\rho_{hi})\lambda_i}.
\label{Lagrange}
\end{eqnarray}
The variables $\delta\rho_{bi}$ and $\delta\rho_{hi}$ also appear in $H_{0}$ and 
$H_{\rm NN}$, but we ignore them.
On integration, linear terms of $\delta\rho_{bi}$ and
$\delta\rho_{hi}$ in $H_{\rm BH}$ are absent as we require the minimal energy condition 
to determine the mean values of $\rho_{bi}$ and $\rho_{hi}$. 
Please see the later discussion.
As we remarked in the above,  
quadratic terms of $\delta\rho_{bi}$ and $\delta\rho_{hi}$ in $H_0$ are partly
incorporated in $H_1$, although  the precise estimation of $V_0$ is
lacking.
For the case with $V>0$, which is discussed in Sec.~VI, the quadratic terms
of the density fluctuations in the repulsion term generate spatially nonlocal terms of 
$\partial_\tau\theta_{bi}$ and $\partial_\tau\theta_{hi}$.
We shall ignore these terms in the practical calculation and therefore we may
underestimate the phase-ordered states.
With this approximation, we have,
\begin{eqnarray}
&&\int d\lambda_{i} d\delta\rho_{bi} d\delta\rho_{hi} \exp\Big[\int d\tau
\Big(-V_{0}\left(\delta\rho_{bi}^2+\delta\rho_{hi}^2\right) \nonumber\\
&&+i\delta\rho_{bi}(\partial_\tau \theta_{bi}+\lambda_{i})+i\delta\rho_{hi}(\partial_{\tau} \theta_{hi}+\lambda_{i})\Big)\Big] \nonumber \\
&&=\int d\lambda_{i} e^{-{1 \over 4V_{0}}\int d\tau\left( (\partial_{\tau} \theta_{bi}+\lambda_{i})^2
+(\partial_{\tau} \theta_{hi}+\lambda_{i})^2\right)},
\label{Lagrange2}
\end{eqnarray}
where we have ignored the terms like $\int d\tau \partial_\tau \theta_{bi}$ ($\int d\tau \partial_\tau \theta_{hi}$)
by the periodic boundary condition for the imaginary time.
The resultant quantity on the right-hand side (RHS) of Eq.(\ref{Lagrange2}) is 
positive-definite,
and therefore the numerical study by the MC simulation can be performed without 
any difficulties.
It should be remarked that the Lagrange multiplier $\lambda_i$ in Eq.(\ref{Lagrange2}) behaves as a gauge field, i.e.,
the RHS of Eq.(\ref{Lagrange2}) is invariant under the following ``gauge transformation",
$\theta_{bi}\rightarrow \theta_{bi}+\alpha_i, \
\theta_{hi}\rightarrow \theta_{hi}+\alpha_i, \ 
\lambda_i \rightarrow \lambda_i-\partial_\tau \alpha_i$.
It is easily shown that all physical quantities
are invariant under the above gauge transformation.
Finally, we have an effective action $S$, with which the partition function is given 
as follows,
\begin{eqnarray}
Z&=&\int [d\theta_{bi}][d\theta_{hi}]e^{-S},\\
\label{effective_Z}
S&=&\int d\tau \biggr(\sum_{i}\frac{1}{4V_{0}}\left( (\partial_{\tau} \theta_{bi}+\lambda_{i})^2
+(\partial_{\tau} \theta_{hi}+\lambda_{i})^2\right)\nonumber\\
&+&J_{1}\sum_{\langle i,j\rangle}\sqrt{\rho_{bi}\rho_{hi}\rho_{bj}\rho_{hj}}\cos(\theta_{i}-\theta_{j})\nonumber\\
&+&J_{2}\sum_{\langle\langle i,j\rangle\rangle}\sqrt{\rho_{bi}\rho_{hi}\rho_{bj}\rho_{hj}}\cos(\theta_{i}-\theta_{j}-\phi_{ij}) \biggl)  \nonumber  \\
&+&V\sum_{\langle i,j \rangle}\rho_{bi}\rho_{bj},
\label{effective_S}
\end{eqnarray}
where $\theta_{i}\equiv\theta_{bi}-\theta_{hi}$.
As the slave particles always appear in the composite $b_ih^\dagger_i$,
the symmetric degrees of freedom $\theta_{i}\equiv\theta_{bi}+\theta_{hi}$ 
decouple, except the first kinetic term of the action $S$ including 
$\partial_\tau \theta_{bi}$ and $\partial_\tau \theta_{hi}$.

%%%%%%%%%%%%%%%%%%%%%%%%%%%%%%%%%%%%%%%%%%%%%%%%%%
\section{Extended Monte-Carlo simulation}\label{eMC}

In the previous section, the effective action $S$ was derived.
For the MC simulation, we introduce a lattice in the imaginary-time $\tau$-direction 
with the lattice spacing $\Delta\tau$.
In order to impose the local constraint Eq.(\ref{constraint}), $\rho_{bi}+\rho_{hi}=1$,
we parameterize $\rho_{bi}$ and $\rho_{hi}$ as 
$\rho_{bi,\ell}=\sin^2(\chi_{i,\ell}), \ \rho_{hi,\ell}=\cos^2(\chi_{i,\ell})$,
where $\chi_{i,\ell}$ is angle variable and $(i,\ell)$ denotes site
in a stacked honeycomb lattice ($\ell$ is imaginary-time index).
Thus, the effective action $S$ becomes a kind of 3D $XY$ model defined on the 
space-time lattice, whereas its coefficients depend on the variational parameters $\{(\rho_{bi,\ell}, \ \rho_{hi,\ell})\}$.
The lattice action and partition function of the lattice model are given as follows:
\begin{eqnarray}
Z_{\rm L}&=&\int\prod^{N_\tau-1}_{\ell=0}
\prod_i [d\chi_{i,\ell}d\theta_{bi,\ell}d\theta_{hi,\ell}d\lambda_{i,\ell}]
e^{-S_{\rm L}},\\
\label{MC_Z} 
S_{\rm L}&=&\sum^{N_\tau-1}_{\ell=0}\Big[\sum_{i}-{1\over 2{V_{0}}_\tau\Delta \tau}
\cos(\theta_{bi,\ell+1}-\theta_{bi,\ell}+\lambda_{i,\ell}) \nonumber \\
&+&\sum_{i}-{1\over 2{V_{0}}_\tau\Delta \tau}
\cos(\theta_{hi,\ell+1}-\theta_{hi,\ell}+\lambda_{i,\ell}) \nonumber \\
&+&{1 \over 2}J_1\Delta\tau\sum_{\langle\langle i,j\rangle\rangle}\sin(2\chi_{i,\ell})
\sin(2\chi_{j,\ell})\cos(\theta_{i,\ell}-\theta_{j,\ell}) \nonumber \\
&+&{1 \over 2}J_2\Delta\tau\sum_{\langle i,j\rangle}\sin(2\chi_{i,\ell})
\sin(2\chi_{j,\ell})\cos(\theta_{i,\ell}-\theta_{j,\ell}-\phi_{ij}) \nonumber \\
&+&V\Delta\tau\sum_{\langle i, j\rangle}
\sin^2(\chi_{i,\ell})\sin^2(\chi_{j,\ell}) \nonumber  \\
&-&\sum_i\ln(\sin(2\chi_{i,\ell}))\Big],
\label{MC_S}
\end{eqnarray}
where $N_\tau$ is the linear system size of the $\tau$-direction and is related
to the temperature ($T$) as $N_\tau\Delta\tau=1/(k_{\rm B}T)$, and all variables 
are periodic in the $\tau$-direction.
It should be remarked here that $\Delta\tau$ is nothing but the inverse
temperature and by changing $\Delta\tau$, $T$ is controlled.
The last term in Eq.(\ref{MC_S}) comes from the change of variables
from $(\rho_{bi,\ell}, \ \rho_{hi,\ell})$ to $\chi_{i,\ell}$.
As we explained above,
as the physical (original) particle $a_i$ is the composite of $b_i$ and $h^\dagger_i$
[Eq.(\ref{slave})], the effective model is invariant under a local gauge
transformation such as 
$
(\theta_{bi,\ell}, \theta_{hi,\ell},\lambda_{i,\ell}) \to
(\theta_{bi,\ell}+\alpha_{i,\ell}, \theta_{hi,\ell}+\alpha_{i,\ell},
\lambda_{i,\ell}-\alpha_{i,\ell+1}+\alpha_{i,\ell})
$
where $\alpha_{i,\ell}$ is an arbitrary parameter ($\alpha_{i,N_\tau+1}=\alpha_{i,1}$).
It seems that the ``gauge field" $\lambda_{i,\ell}$ can be eliminated by the gauge fixing,
but this is not the case.
After the gauge fixing, there remains one degrees of freedom per site $i$, i.e., so-called
zero mode, $\sum_{\ell=1}^{N_\tau}\lambda_{i,\ell}$.
In the MC simulation, we remain the gauge field $\lambda_{i,\ell=1}$ 
as MC variables whereas we put the others $\lambda_{i,\ell\neq 1}=0$.

The effective action in the path-integral formalism includes both the
variational parameters 
$\{(\rho_{bi,\ell}, \ \rho_{hi,\ell})\} (\rightarrow \{\chi_{i,\ell}\})$
and the dynamical phase variables, $\{\theta_{bi}\}$ and $\{\theta_{hi}\}$.
We determines the variational variables $\{(\rho_{bi,\ell}, \ \rho_{hi,\ell})\}$
by the minimum-energy condition by using MC methods.
In the practical calculation of Eq.(\ref{MC_S}), we treat $\{(\rho_{bi,\ell}, \ \rho_{hi,\ell})\}$
as slow variables in the MC local-update, keeping the mean densities constant.
As the effective action $S_{\rm L}$ in Eq.(\ref{MC_S}) is real and bounded from below,
there exist no difficulties in performing MC simulations.
In the following sections, we shall show the numerical results and discuss  the physical meaning of them.

%%%%%%%%%%%%%%%%%%%%%%%%%%%%%%%%%%%%%%%%%%%%%%%%%%
\section{Numerical results for $V=\phi=0$ case}\label{results}

In this section and subsequent sections, we shall show the results obtained 
by the MC simulation.
The effective model is defined by Eq.(\ref{MC_S}) and we employ the standard 
Metropolis algorithm with the local updates \cite{Metropolis}.
For the local update of the phase degrees of freedom $\theta$, random
variables $\Delta \theta$ used for generating a candidate of a new variable
$\theta_{new}=\theta_{old}+\Delta \theta$ was chosen in the range 
$|\Delta \theta| \leq \pi/6$.
Furthermore in this study, the local average densities are
also variational parameters and are parameterized by the angle variables 
$\{\chi_{i,\ell}\}$.
Since the local average densities are slow variables, the range of random variables
$\Delta \chi_{i,\ell}$ are restricted as $|\Delta \chi_{i,\ell}| \leq \pi/60$.
The typical sweep for the thermalization is 100 000 and for the measurement is 
(40 000)$\times$(10 samples).
The typical acceptance ratio was 40\%$\sim$50\%, and errors were estimated from 10 samples by the jackknife method \cite{Jackknife}.

%%%%%%%% FIG1
\begin{figure}[h]
%\vspace{0.5cm}
\begin{center}
\includegraphics[width=9cm]{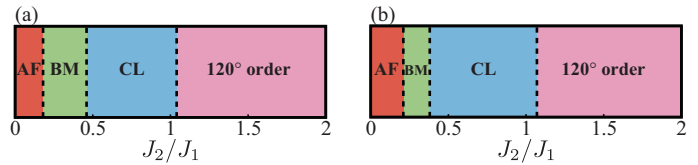}
\end{center}
\caption{(Color online) Phase diagrams of the Bose-Hubbard model on the 
honeycomb lattice with $V=\phi=0$.
(a) $V_0=5$, (b) $V_0=0.5$.
There are four phases, antiferromagnetic (AF), Bose metal (BM), 
colinear (CL) and $120^o$-order state.
Lattice size is small, $(L_x,L_y)=(3,4)$.}
\label{PD1}
\end{figure}
%%%%%%%%%%%%%%
%%%%%%%% FIG2
\begin{figure}[h]
\begin{center}
\includegraphics[width=8cm]{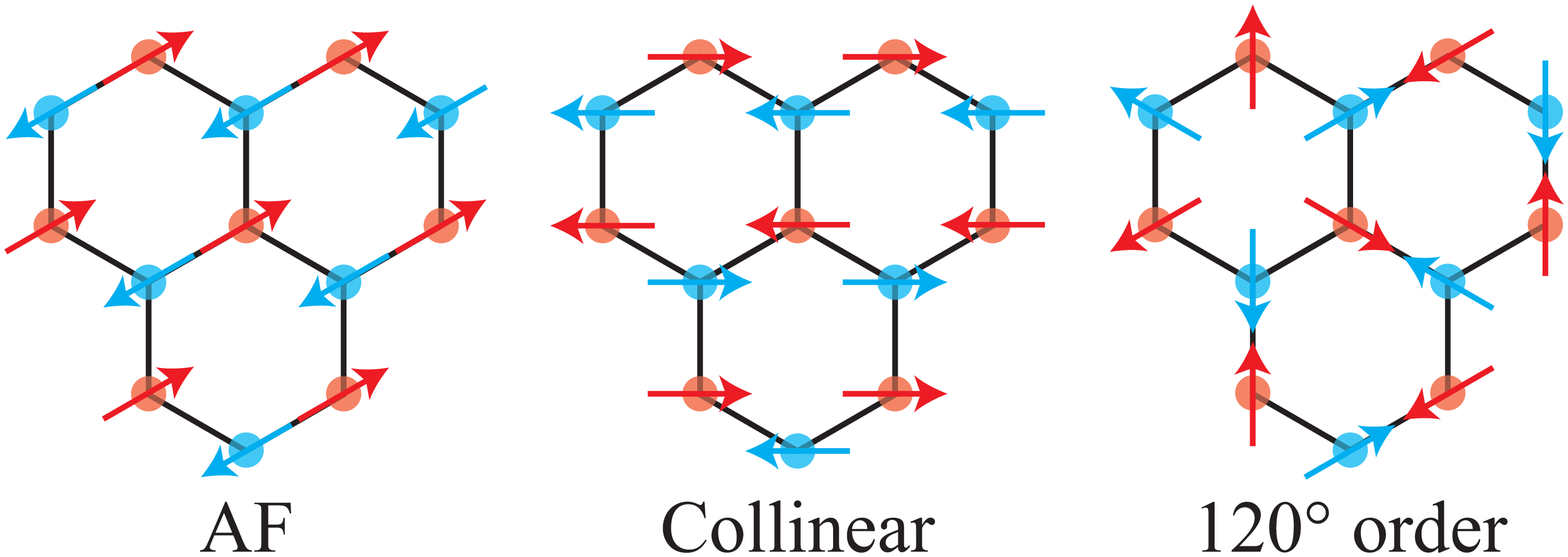}
\end{center}
\caption{(Color online) Spin (phase $\theta_i$) configurations of the three ordered states,
antiferromagnetic (AF), colinear (CL) and $120^o$-order phase in the phase diagram
in Fig.~\ref{PD1}.}
\label{spin1}
\end{figure}
%%%%%%%%%%%%%%

%%%%%%%%%%%%%%%%%%%%%%%%%%%%%%%%%%%%%%%%%%%%%%%%%%%%%%%%%%%%%%%%%%%
\subsection{Low-temperature phase diagrams}

We first show the phase diagrams as a function of the dimensionless parameter
$J_{2}/J_{1}$ for the case of $V=\phi=0$ \cite{Bishop}. 
In the practical calculation, we put $J_1=10$ and $\Delta\tau=1$.
See Fig.~\ref{PD1} for the phase diagrams for $V_0=5$ and $V_0=0.5$.
For the case of the system size $(L_x,L_y)=(3,4)$ and $N_\tau=8$, there are four phases,
i.e., AF, BM, collinear (CL), and $120^o$-order state.
There exist no qualitative differences between the phase diagrams of the $V_0=5$ 
and $V_0=0.5$ cases.

%%%%%%%% FIG3
\begin{figure}[h]
%\vspace{0.5cm}
\begin{center}
\includegraphics[width=3cm]{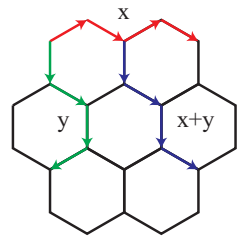}
\end{center}
\caption{(Color online) $x$, $y$ and $x+y$-directions on the honeycomb lattice.
}
\label{HClattice}
\end{figure}
%%%%%%%%%%%%%%

To clarify the phase diagrams, we calculated various physical quantities.
Phase boundaries were determined by calculating the ``internal energy" $E$
and the ``specific heat" $C$, which are defined as
\begin{eqnarray}
E={\langle S_{\rm L} \rangle \over N}, \;\;\;
C={\langle (S_{\rm L}-\langle S_{\rm L} \rangle)^2 \rangle \over N},
\label{EC}
\end{eqnarray}
where $N=N_\tau N_{\rm 2D}$ is the total number of sites in the stacked 
honeycomb lattice and we employ the periodic boundary condition.
We also calculate correlation functions on the honeycomb lattice in the
$x$, $y$, and $x+y$ directions (see Fig.~\ref{HClattice}), which are defined as follows,
\begin{eqnarray}
&&G_{\rm x(y)(x+y)}(r)={1 \over N_{\rm 2D}}
\sum_i\langle \cos(\theta_{i+r}-\theta_i)\rangle,
\label{corr}
\end{eqnarray}
where the site $i+r$ denotes the sites with distance $r$ from the site $i$ 
in the $x, \ y$ and $x+y$ directions in the honeycomb lattice, respectively.

For the phases except the BM, the order
parameter, $\langle a_i \rangle$, has a coherent phase, 
as shown by the calculated correlation functions.
Spin (phase of $\langle a_i\rangle=\theta_i$) configurations for the phases in the phase diagram in
Fig.~\ref{PD1} are depicted in Fig.~\ref{spin1}.
From Fig.~\ref{spin1}, we identified the AF, CL, and 120$^o$-order phase.
As we show shortly for a larger system, the BM has only a short-range correlation of 
$\langle a_i \rangle$.

There are results of the exact diagonalization for the system with the 
size  $(L_x,L_y)=(3,4)$ ($N_{\rm 2D}=24$).
Qualitatively the same phase diagram with those in Fig.~\ref{PD1} was obtained.
Critical values of $J_2/J_1$ of the exact diagonalization at which the phase transitions
take place are very close to those obtained in this work,
in particular, the result with $V_0=0.5$.

%%%%%%%% FIG4 
\begin{figure}[h]
%\vspace{0.5cm}
\begin{center}
\includegraphics[width=9cm]{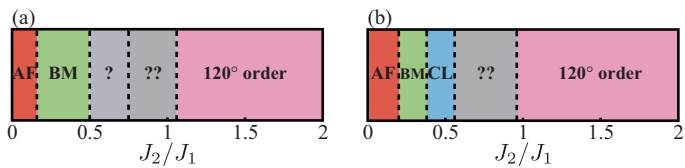}
\end{center}
\caption{(Color online) Phase diagrams of the Bose-Hubbard model on the 
honeycomb lattice with $V=\phi=0$.
(a) $V_0=5$, (b) $V_0=0.5$.
There are unidentified phases besides antiferromagnetic (AF), Bose metal (BM), 
colinear (CL) and phase with the $120^o$ order.
Lattice size, $(L_x,L_y)=(6,6)$.}
\label{PD2}
\end{figure}
%%%%%%%%%%%%%%

As the system has the strong frustrations for the case of $\phi=0$,
it is important to see if the phase diagram depends on the system size.
In order to see this, we studied the system with $(L_x,L_y)=(6,6)$.
No exact diaganalization results are available for this system size.
Obtained phase diagrams are shown in Fig.~\ref{PD2}.
Compared with the previous result of $(L_x,L_y)=(3,4)$, there are additional
phases whose spin (i.e., phase $\theta_i$) configuration cannot be depicted globally.

%%%%%%%% FIG5 
\begin{figure}[h]
\vspace{0.5cm}
\begin{center}
\includegraphics[width=9cm]{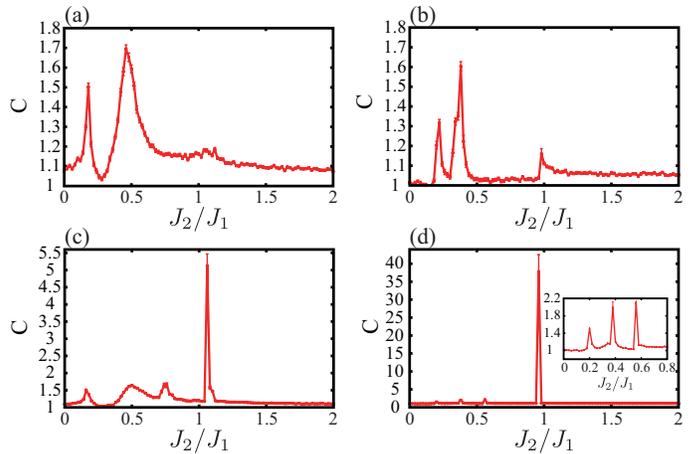}
\end{center}
\caption{(Color online) Specific heat $C$ for the case of $V=\phi=0$.
(a) $(L_x,L_y)=(3,4)$ and $V_0=5$, (b)  $(L_x,L_y)=(3,4)$ and $V_0=0.5$,
(c) $(L_x,L_y)=(6,6)$ and $V_0=5$, (d)  $(L_x,L_y)=(6,6)$ and $V_0=0.5$.
Sharp peaks in the system with  $(L_x,L_y)=(6,6)$ indicate that the transition
is of first order.
This is verified by the measurement of the internal energy $E$.
}
\label{SpH}
\end{figure}
%%%%%%%%%%%%%%

%%%%%%%% FIG6 
\begin{figure*}[t]
%\vspace{0.5cm}
\begin{center}
\includegraphics[width=12cm]{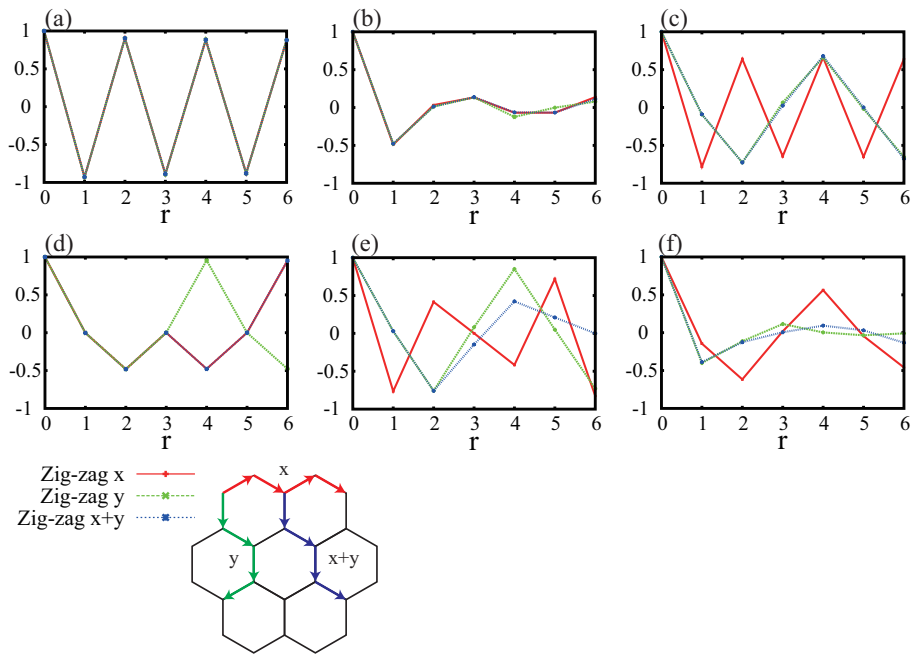}
\end{center}
\caption{(Color online) Correlation functions in Eq.(\ref{corr}) for the
phases in Fig.~\ref{PD2}.
(a) AF, (b) BM, (c) CL, (d) 120$^o$-order state, and (e) and (f) are phases 
whose existence is indicated by the calculation of the specific heat.
}
\label{CF}
\end{figure*}
%%%%%%%%%%%%%%

Calculations of the specific heat used to obtain the phase diagrams
in Figs.~\ref{PD1} and \ref{PD2} are shown in Fig.~\ref{SpH}.
As the system size is getting larger, not only the peaks of $C$ get sharper
but also other peaks appear.
Compared with systems without frustrations studied in previous works, 
the specific heat $C$ has very strong system-size dependence in the present case.

To identify the phases, we calculated the correlation functions in Eq.(\ref{corr}).
The results are shown in Fig.~\ref{CF}.
We verified that all the numerical results are quite stable even for the 
unidentified phases in the phase diagram in Fig.~\ref{PD2}.
Phases (a), (c) and (d) are the AF, CL and 120$^o$ phases, respectively,
and the phase (b) is the BM without long-range correlations.
The calculation of the specific heat indicates the existence of
the phases (e) and (f).
The phase (e) and (f) have rather clear spin correlations as shown in Fig.~\ref{CF},
but it is difficult to depict global configurations of the phase of $\langle a_i \rangle$.

%%%%%%%% FIG7 
\begin{figure}[h]
\begin{center}
\includegraphics[width=8cm]{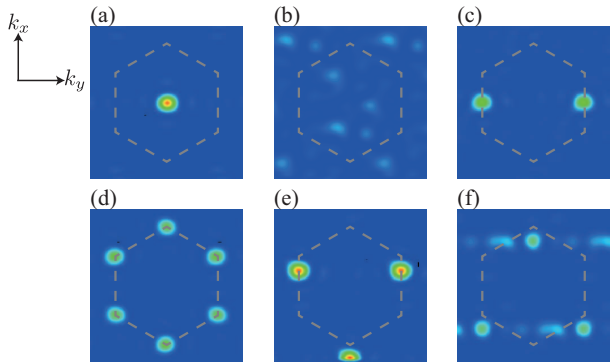}
\end{center}
\caption{(Color online) Density distribution in the momentum space
$n({\bf k})$ defined by Eq.(\ref{density}) for the phases in Fig.~\ref{PD2}.
(a) AF, (b) BM, (c) CL, (d) 120$^o$-order state, and (e) and (f) are phases 
whose existence is indicated by the calculation of the specific heat.
The dotted line denotes the boundary of the Brillouin zone.
}
\label{particleD}
\end{figure}
%%%%%%%%%%%%%%

It is sometimes useful to see the particle density in the wave-vector space,
which is defined as,
\begin{eqnarray}
n({\bf k})&=&\langle a^\dagger({\bf k})a({\bf k})\rangle  \nonumber \\
&=&\sum_{i,j}e^{-{\bf k}\cdot ({\bf r}_i-{\bf r}_j)}\langle a^\dagger_ia_j\rangle,
\label{density}
\end{eqnarray}
where ${\bf r}_i$ and ${\bf r}_j$ are lattice vectors corresponding to sites $i$ and $j$
of the honeycomb lattice, respectively.
We show the calculations of $n({\bf k})$ in Fig.~\ref{particleD}.
In the AF, CL and 120$^o$-order phases, particles condensate at some specific
momenta consistent with the correlation function, whereas in the BM 
no clear pattern can be seen.
On the other hand for the phases (e) and (f), the particle density has moderate
but rather clear peaks at three spots although 
their locations are incommensurate with the Brillouin zone.

It is quite interesting to see if the BM is gapless or gapful.
To study this problem, we notice that a change of the parameter $\Delta\tau$
corresponds to a change of the system temperature $T$.
From this fact, we can measure the $T$-dependence of the specific
heat $C$.
Furthermore, if signals of phase transition do not appear as $T$ is increased, 
we conclude that the BM has no LROs.

%%%%%%%%%%%%%%%%%%%%%%%%%%%%%%%%%%%%%%%%%%%%%%%%%%
\subsection{Finite-temperature phase transitions for $V=\phi=0$ case}

%%%%%%%% FIG8 
\begin{figure}[h]
%\vspace{0.5cm}
\begin{center}
\includegraphics[width=8cm]{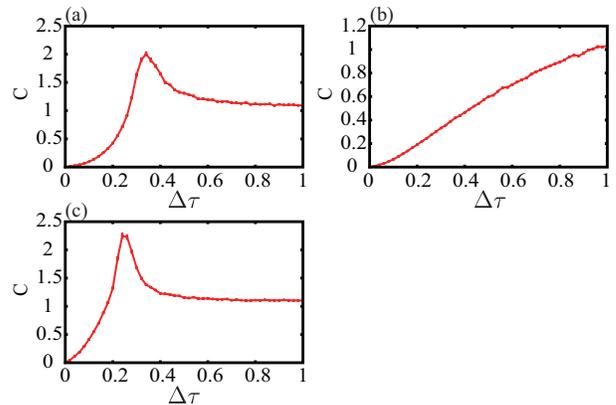}
\end{center}
\caption{(Color online) Finite-$T$ phase transitions for the AF [(a)], 
BM [(b)], and 120$^o$-order state [(c)] phases, respectively.
The thermal specific heat $C_T$ exhibits a sharp peak for the AF and 
the 120$^o$-order phase, whereas there is no signal of the phase transition 
for the BM.
This result indicates that the BM does not have any long-range orders.
$C_T$ for the AF and the 120$^o$-order phase approaches to a constant close to unity
at low $T$ limit.
$V_0=5$.
}
\label{finiteT}
\end{figure}
%%%%%%%%%%%%%%

In this section, we shall study finite-$T$ effects on the phases
observed in the previous subsection \cite{Mori}.
In particular, it is interesting to see if a phase transition takes place or not
in the BM, i.e., the existence of a finite-$T$ phase transition means that
the BM phase has a certain order at low (vanishing) $T$, or vice versa.
On the other hand, we expect that a finite-$T$ phase transition takes place
at a certain critical $T_c$ for the AF and 120$^o$-order states.

As we explained previously, $N_\tau\Delta\tau=1/(k_{\rm B}T)$, and 
therefore a decrease of $\Delta\tau$ corresponds to an increase of $T$.
The finite-$T$ system is described by the effective action $S_{\rm L}$ in 
Eq.(\ref{MC_S}) with $\Delta\tau<1$.
It should be remarked here that the present numerical parameters such as
$J_1\Delta\tau=10$ and $N_\tau=8$ corresponds to 
$k_{\rm B}T=J_1/80$, which means a very low $T$.
We study the system with the lattice size $(L_x,L_y)=(6,6)$, and 
focus on the AF, BM, and 120$^o$-order state.

Numerical result of {\em the thermal specific heat} $C_T$ is given in Fig.~\ref{finiteT},
where the thermal specific heat $C_T$ is defined as follows,
\begin{eqnarray}
C_T&=&{1 \over N_{\rm 2D}}\Big[\langle (H_\beta)^2\rangle
-\langle H_\beta \rangle^2\Big],  \nonumber \\
H_\beta &\equiv &
\sum^{N_\tau-1}_{\ell=0}\Big[
{1 \over 2}J_1\Delta\tau\sum_{\langle\langle i,j\rangle\rangle}\sin(2\chi_{i,\ell})
\sin(2\chi_{j,\ell})\cos(\theta_{i,\ell}-\theta_{j,\ell}) \nonumber \\
&+&{1 \over 2}J_2\Delta\tau\sum_{\langle i,j\rangle}\sin(2\chi_{i,\ell})
\sin(2\chi_{j,\ell})\cos(\theta_{i,\ell}-\theta_{j,\ell}) \nonumber \\
&+&V\Delta\tau\sum_{\langle i, j\rangle}
\sin^2(\chi_{i,\ell})\sin^2(\chi_{j,\ell}) \nonumber  \\
&-&\sum_i\ln(\sin(2\chi_{i,\ell}))\Big].
\label{CT}
\end{eqnarray}
It is obvious that $C_T$ exhibits a sharp peak for the AF and 120$^o$-order state,
whereas no peaks for the BM.
This result means that the BM does not have any long-range orders.
We verified that the orders of the AF and 120$^o$-order state are destroyed
at $T_c$ identified by $C_T$.

At low $T<T_c$ ($\Delta\tau>\Delta\tau_c$), $C_T$ for the AF and 120$^o$-order
state has a constant value close to unity.
This behavior indicates that a stable quasi-excitation exists that is nothing but 
a Nambu-Goldstone mode appearing as a result of the spontaneous U(1)
symmetry breaking.
On the other hand for the BM, $C_T$ increases as $T$ decreases.
This implies that excitations in the BM are {\em not} simple gapless quasi-particles
and a strongly-correlated (strongly-frustrated) state forms in the BM.

%%%%%%%%%%%%%%%%%%%%%%%%%%%%%%%%%%%%%%%%%%%%%%%%%

\section{Phase diagram in ($J_2/J_1$-$\phi$) plane}

%%%%%%%% FIG9 
\begin{figure}[h]
\begin{center}
\includegraphics[width=6.5cm]{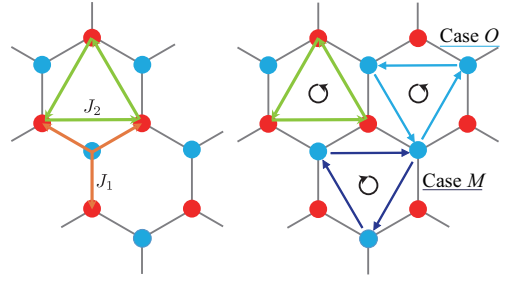}
\end{center}
\caption{(Color online) Hopping terms in the BHHM.
The NNN hopping amplitudes with an arrow are $J_2e^{i\phi}$.
Case $O$ ($M$) refers to the original (modified) model.
}
\label{originalmodify}
\end{figure}
%%%%%%%%%%%%%%

In this section, we study the phase diagram of the BHHM
with nonvanishing $\phi$ (the phase of the NN hopping), whereas 
we keep the NN repulsion $V=0$.
As the case of $\phi=0$ is the most frustrated system, it is expected that
turning on $\phi$ makes the system more tractable and stabilizes the 
ground-state.
There are two possible ways to introduce the phase $\phi_{ij}$ in the NN hopping
as depicted in Fig.~\ref{originalmodify}, one of which is called original Haldane model 
and the other 
is called modified Haldane model \cite{modified}.
As we show, these two models can have different phase diagrams as in the
ferromagnetic NN coupling cases studied in the previous paper \cite{Nakafuji}.

%%%%%%%% FIG10 
\begin{figure}[h]
%\vspace{0.5cm}
\begin{center}
\includegraphics[width=4cm]{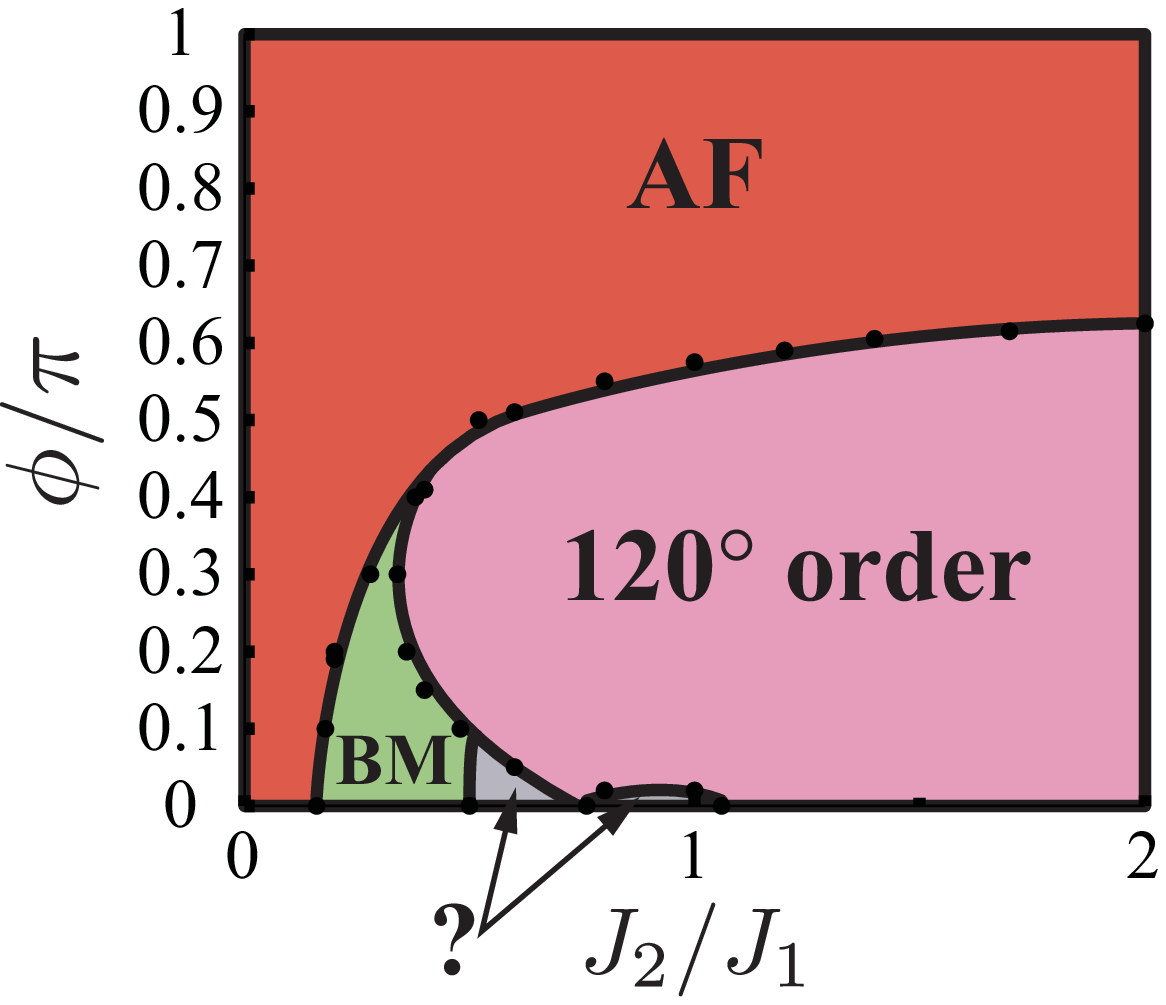}
\includegraphics[width=4cm]{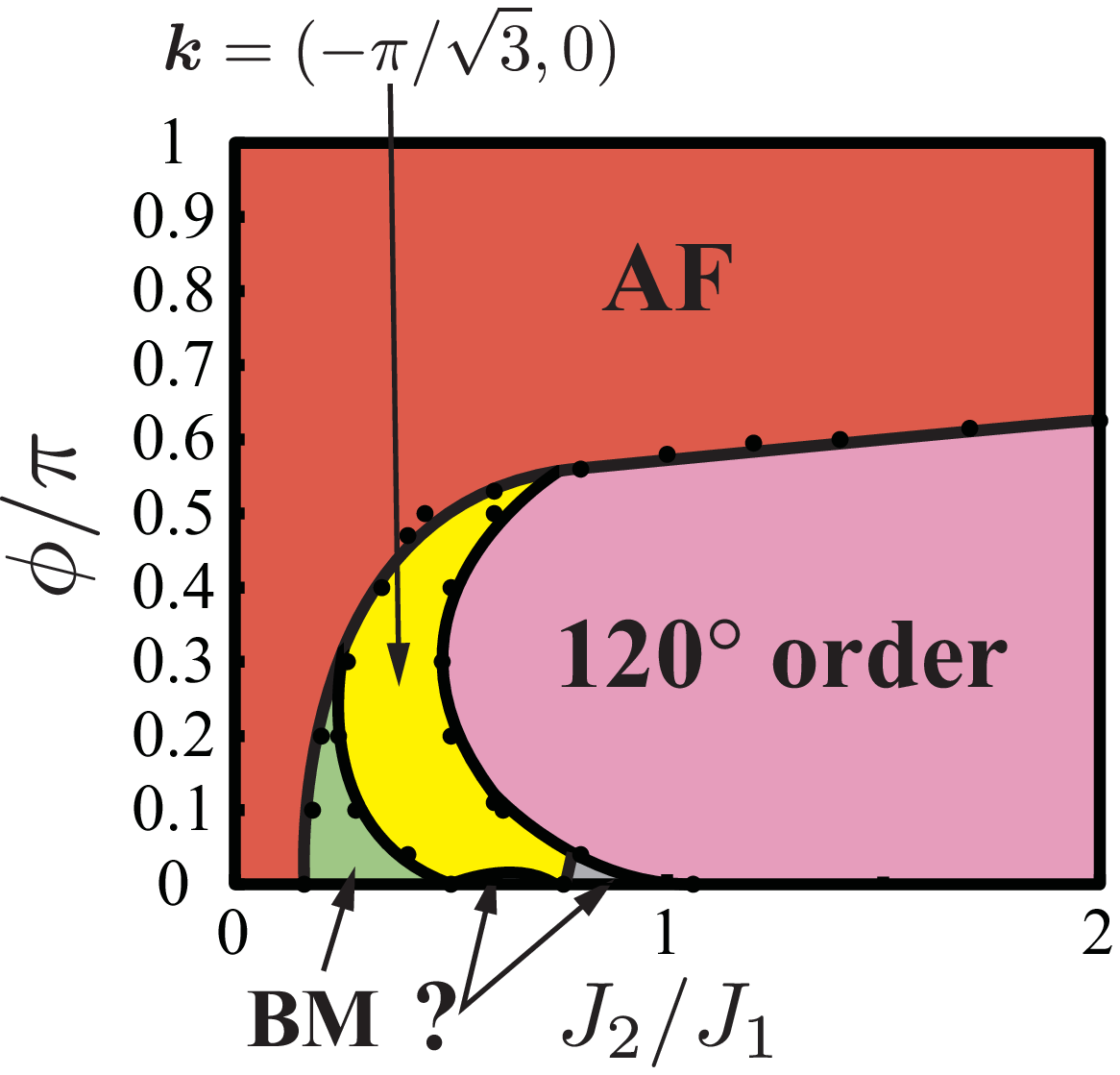}
\end{center}
\caption{(Color online) Phase diagrams of the original (left) and modified (right)
BHHMs.
Finite $\phi$ makes the system less frustrated, and clear phase boundaries are
obtained.
$V_0=5$ and the system size $(L_x,L_y)=(6,6)$.
}
\label{PD3}
\end{figure}
%%%%%%%%%%%%%%

By using the same extended MC simulations, we obtained the phase diagrams of 
the original and modified BHHMs as in Fig.~\ref{PD3}.
Even for very small $\phi$, the phase boundaries are obtained rather clearly.
Calculations of the internal energy and specific heat indicate that the phase transition
AF $\leftrightarrow$ 120$^o$-order state is of first order, whereas both
AF $\leftrightarrow$ BM and 120$^o$ $\leftrightarrow$ BM transitions are 
continuous ones.
This result again implies that the BM does not have any long-range orders.
In the experimental set up to realize the Bose-Haldane model on the honeycomb
lattice, it is expected that the value of $\phi$ can be a controllable parameter.
Careful study on the phase diagram by experiments may shed light on the physical
properties of the phases with $\phi=0$ and related AF magnets on the honeycomb
lattice.

%%%%%%% FIG11 
\begin{figure}[h]
\begin{center}
\includegraphics[width=9cm]{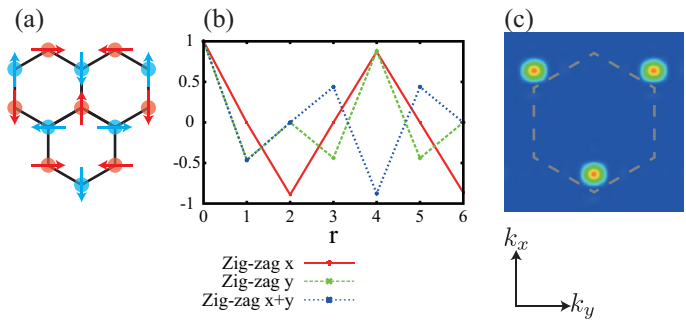}
\end{center}
\caption{(Color online) Properties of the ${\bf k}=(-\pi/\sqrt{3}, 0)$ phase.
(a) spin configuration, (b) correlation function, and (c) density distribution
in the momentum space.
}
\label{K-state}
\end{figure}
%%%%%%%%%%%%%%

In the phase diagram of the modified BHHM, there appears another
ordered phase that we call ${\bf k}=(-\pi/\sqrt{3}, 0)$ phase.
Details are shown in Fig.~\ref{K-state}.
Phase closely related to this
was recently observed for a frustrated Heisenberg model on 
the honeycomb lattice \cite{Ciolo}.
In the present case, the reduction of the frustration by a finite $\phi$ makes
this phase stable for a rather large region of the phase diagram.
Angles between the NN spins and also NNN spins are either $\pi/2$ or $\pi$.
The particle distribution functions $n({\bf k})$ in Fig.~\ref{particleD}(e)  and
Fig.~\ref{K-state}(c) indicate that the phase in Fig.~\ref{particleD}(e) 
and the ${\bf k}=(-\pi/\sqrt{3}, 0)$ phase
have some similarity although there exists a phase boundary between them 
as shown in the phase diagram of Fig.~\ref{PD3}.

%%%%%%%%%%%%%%%%%%%%%%%%%%%%%%%%%%%%%%%%%%%%%%%%%%
\section{Phase diagram with NN interaction}

In this section, we study the system in which the NN interaction $H_{\rm NN}$
in Eq.~(\ref{H0}) exists \cite{Rigol}.
As mentioned previously, this term corresponds to the AF-spin coupling in 
the $z$-component such as $V\sum_{\langle i,j\rangle}S^z_iS^z_j$.
The previous study on some related models shows that the inclusion of the NN
interaction $H_{\rm NN}$ makes the system stable as the frustration in the $XY$
component is weakened by the existence of this term \cite{Nakafuji}.
For sufficiently large $V$, it is expected that the charge-density order appears,
which corresponds to the AF order in the $z$-component of spin.

%%%%%%% FIG12 
\begin{figure}[h]
%\vspace{0.5cm}
\begin{center}
\includegraphics[width=8cm]{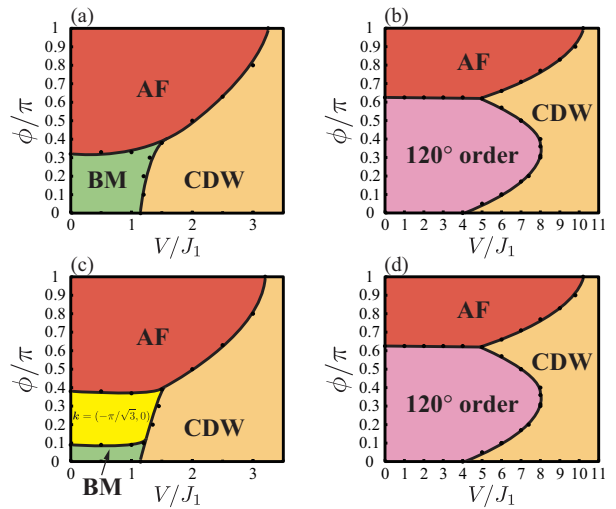}
\end{center}
\caption{(Color online) Phase diagrams of the BHHM with the NN interaction $V>0$.
$J_1=10$ and $V_0=5$. 
(a) original BHHM with $J_2=3$, (b) original BHHM with $J_2=20$, 
(c) modified BHHM with $J_2=3$, (d) modified BHHM with $J_2=20$.
For $J_2=20$, the phase diagrams of the original and modified BHHM
are almost the same as the NNN hopping dominates over the NN hopping.
The system is at half filling with $\mu/V=1.5$.
}
\label{PD4}
\end{figure}
%%%%%%%%%%%%%%

%%%%%%% FIG13 
\begin{figure}[h]
%\vspace{0.5cm}
\begin{center}
\includegraphics[width=8cm]{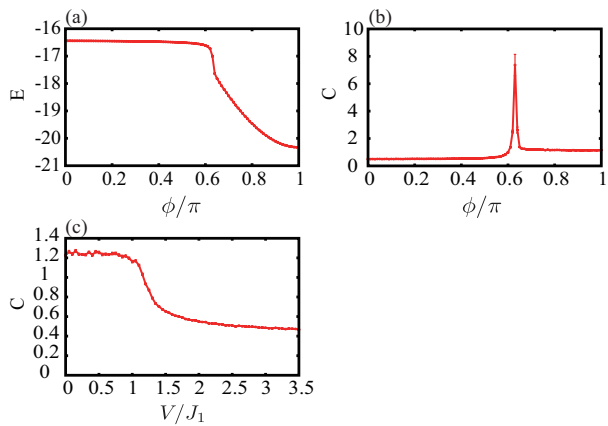}
\end{center}
\caption{(Color online) Internal energy $E$ and specific heat $C$ measured
in the phases in Fig.~\ref{PD4}.
$V_0=5$. 
(a) $E$ as a function of $\phi$ in the original BHHM with $J_1=10,J_2=3$ and $V=25$,
(b) $C$ as a function of $\phi$ in the original BHHM with $J_1=10,J_2=3$ and $V=25$
(c) $C$ as a function of $V/J_1$ of the original BHHM with $J_2=3.0$.
Phase transition between the AF and CDW is of first order, and that between
the BM and CDW is of second order.
The system is at half filling with $\mu/V=1.5$, where $\mu$ is the chemical potential 
of the BHHM.
}
\label{EC2}
\end{figure}
%%%%%%%%%%%%%%

We investigated the system with $\phi=0$ and $V>0$ by the extended MC
simulations as before and obtained 
the phase diagrams shown in Fig.~\ref{PD4}.
The system is at half filling with $\mu/V=1.5$, i.e., $\langle \sum_iS_i^z \rangle=0$,
and we put $J_1=10$ and $\Delta\tau=1$ as before.
For small $J_2$ and $\phi$, the BM still exists as in the pure $XY$ case.
As $V$ is getting large, the state with the charge-density wave (CDW) forms.
The phase transition AF $\leftrightarrow$ CDW and 
120$^o$-order state $\leftrightarrow$ CDW are both first-order phase transitions,
whereas the transition between the BM and CDW is a continuous one.
The internal energy $E$ and specific heat $C$ are shown in Fig.~\ref{EC2}.
At the AF $\leftrightarrow$ CDW phase transition, $E$ exhibits 
a step-wise behavior although hysteresis is not observed.
At the critical point, $C$ has a very large and steep peak.
On the other hand, $C$ shows a step-wise behavior at the BM $\leftrightarrow$ CDW
transition.
Although $C$ has no sharp peak, this anomalous behavior of $C$ indicates that 
a second-order phase transition or a crossover takes place between the BM
and CDW.
Again, this result indicates that there are no LROs in the BM.

For the AF XY spin model on the honeycomb lattice of the cylinder geometry,
the density-matrix renormalization group study showed that in the intermediate
parameter region of $J_2/J_1$, unexpected AF order in the $z$-direction 
forms \cite{DMRG}.
In the present study on the hard-core BHHM, the above parameter region
corresponds to the BM in Fig.~\ref{PD4} (a) with $V,\phi \sim0$.
We measured the density correlation in the BM near the phase boundary 
to the CDW and found a short-range correlation as it is expected.
Then, it is interesting to study the BHHM on the honeycomb lattice of the
cylinder geometry and to see if such a density order persists in the deep BM region.
This problem will be studied and results will be published in the near future.

%%%%%%%%%%%%%%%%%%%%%%%%%%%%%%%%%%%%%%%%%%%%%%%%%%
\section{Conclusion}

In this paper, we studied the hard-core BHHM in which the frustration
caused by the NN and NNN hoppings exists.
Strength of the frustration is controlled by the phase $\phi$ of the NNN
hopping.
This model is closely related with the spin-1/2 AF magnets and it is expected
that a state without any long-range orders exists in a moderate parameter
region of the phase diagram.

We first considered the case with $\phi=0$ and the vanishing NN repulsion, 
i.e., the most frustrated case.
The extended MC simulation shows that the AF, CL and 120$^o$-order
state form as the NNN amplitude increases, while there appears the state that
we call the BM between the AF and CL.
Correlation functions and the order of the phase transition indicate that
the BM has no LROs.
We also studied the finite-$T$ phase diagram and found that no transitions
take place from the BM as $T$ is increased.
Therefore we concluded that the BM is a featureless state.
On the other hand, all the above mentioned ordered states transit to disordered states through the second-order phase transitions.

Results of the MC simulations show a strong system-size dependence of the
phase diagram.
The results for the $(L_x,L_y)=(3,4)$ system (a small system) are in good agreement
with the results obtained by the exact diagonalization for the same system 
size \cite{varney}.
On the other hand, the larger system with $(L_x,L_y)=(6,6)$ has a slightly different
phase diagram in which additional unidentified phases appear.
Investigation of these states is a future problem.
[No results of the exact diagonalization are available for the $(L_x,L_y)=(6,6)$ case.]

Next we studied the phase diagram in the $(J_2/J_1-\phi)$ plane.
There are two types of the BHHM named the original and modified BHHMs,
respectively.
As increasing the value of $\phi$, stable states and phase boundaries appear.
Besides the ordered states in the $\phi=0$ case, there appear another 
ordered state that we call ${\bf k}=(-\pi/\sqrt{3},0)$ state.
Finally, we examined the effect of the NN repulsion.
Inclusion of the NN repulsion, which corresponds to the AF coupling 
$V\sum_{\langle i,j \rangle}S^z_iS^z_j$, stabilizes the frustration.
As $V$ is increased, the CDW forms that corresponds to the Ising-type AF
configuration in the $z$-component.

We obtained the ``multi-dimensional phase diagram" in this work.
The result suggests feasible experiments that quantum simulate the BHHM 
with cold atomic gases on the optical lattice.
We expect that these quantum simulation clarifies the physical nature of the
BM as well as the unidentified states observed in this study.
This must shed light on the physical nature of the spin liquid in the AF
magnets on the honeycomb lattice.  

Finally, recently some related spin and boson models were analytically studied
by using the Chern-Simon gauge theory.
There, dynamical variables are described by using fermions and 
various phase diagrams were obtained \cite{CS}.
It is interesting and also important to extend the present numerical study 
to these models and clarify  the relationship to the fermionic degrees of freedom.

%%%%%%%%%%%%%%%%%%%%%%%%%%%%%%%%%%%%%%%%%%%%%%%%%%
\bigskip

\acknowledgments
This work was partially supported by Grant-in-Aid
for Scientific Research from Japan Society for the 
Promotion of Science under Grant No.26400246.

%%%%%%%%%%%%%%%%%%%%%%%%%%%%%%%%%%%%%%%%%%%%%%%%%%


\begin{thebibliography}{99}
\bibitem{slaveMC}As we pointed out in the previous paper \cite{Nakafuji}, 
the coherent 
states $|\varphi_{b}\rangle, \ |\varphi_{h}\rangle$ contain multi-particle states 
even if they satisfy the local constraint such as 
$|\varphi_b|^2+|\varphi_h|^2=1$.
In most of cases, this fact does not matter to critical behaviors of many-body
quantum systems if there are no frustrations.
The term $H_1$ [Eq.(\ref{H1})] suppresses the multi-particle states in the 
coherent state path integrals and has relevant physical meanings for
systems with frustrations as in the present case.
\end{thebibliography}
\end{document}